\documentclass[12pt, onecolumn, draftclsnofoot]{IEEEtran}

\usepackage[noadjust]{cite}
\usepackage{enumerate}
\usepackage{amssymb, amsmath}

\newtheorem{theorem}{Theorem}[section]
\newtheorem{lemma}[theorem]{Lemma}
\newtheorem{corollary}[theorem]{Corollary}
\newtheorem{definition}[theorem]{Definition}
\newtheorem{construction}[theorem]{Construction}
\newenvironment{proofnote}{\par{\itshape Note added in proof:}}{}

\newcommand{\beqn}{\begin{equation}}
\newcommand{\eeqn}{\end{equation}}
\newcommand{\beq}{\begin{equation*}}
\newcommand{\eeq}{\end{equation*}}
\newcommand{\Z}{\mathbb{Z}}
\newcommand{\Cn}{\mathbb{C}}
\newcommand{\B}{\mathcal{B}}
\newcommand{\C}{\mathcal{C}}
\newcommand{\Q}{\mathcal{Q}}

\newcommand{\DG}{\mathcal{DG}}
\newcommand{\K}{\mathcal{K}}
\newcommand{\M}{\mathcal{M}}

\renewcommand{\d}{{\rm d}}
\renewcommand{\Re}{{\rm Re}}
\renewcommand{\Im}{{\rm Im}}
\DeclareMathOperator{\tr}{tr}
\DeclareMathOperator{\rk}{rk}
\DeclareMathOperator{\rad}{rad}
\DeclareMathOperator{\PAPR}{PAPR}
\DeclareMathOperator{\RM}{RM}
\DeclareMathOperator{\ZRM}{ZRM}

\newcommand{\bone}{\boldsymbol{1}}

\begin{document}
\date{}

\title{Quaternary Constant-Amplitude Codes for \\Multicode CDMA}
\author{\large Kai-Uwe Schmidt\thanks{Kai-Uwe Schmidt was with the Communications Laboratory, Dresden University of Technology, 01062 Dresden, Germany. He is now with the Department of Mathematics, Simon Fraser University, 8888 University Drive, Burnaby, BC V5A 1S6, Canada, e-mail: kuschmidt@sfu.ca}\thanks{This work was supported by the Deutsche Forschungsgemeinschaft (DFG) under grant FI 470/8-1.}}%

\date{\large \today}
\maketitle


\begin{abstract}
A constant-amplitude code is a code that reduces the peak-to-average power ratio (PAPR) in multicode code-division multiple access (MC-CDMA) systems to the favorable value $1$. In this paper quaternary constant-amplitude codes (codes over $\Z_4$) of length $2^m$ with error-correction capabilities are studied. These codes exist for every positive integer $m$, while binary constant-amplitude codes cannot exist if $m$ is odd. Every word of such a code corresponds to a function from the binary $m$-tuples to $\Z_4$ having the bent property, i.e., its Fourier transform has magnitudes $2^{m/2}$. Several constructions of such functions are presented, which are exploited in connection with algebraic codes over $\Z_4$ (in particular quaternary Reed--Muller, Kerdock, and Delsarte--Goethals codes) to construct families of quaternary constant-amplitude codes. Mappings from binary to quaternary constant-amplitude codes are presented as well.
\end{abstract}
 

\section{Introduction}

\PARstart{M}{ulticode} code-division multiple access (MC-CDMA) is a simple scheme to implement rate adaption in CDMA systems \cite{I1995}. The basic idea is to assign additional spreading sequences to a user who wishes to transmit with a higher data rate. In order to avoid self-interference, the user commonly employs $n=2^m$ orthogonal spreading sequences, which can be viewed as the rows of a Hadamard matrix of order $n$. Thus the data rate of this user is $n$ times that in a conventional CDMA system. Typical values of $m$ are $2$ to $6$. The principal drawback of this technique is that the transmitted signals can have a high peak-to-average power ratio (PAPR). That is, the peak transmit power can be much larger than the average transmit power. Thus the efficiency of analog devices in the transmission chain, such as the power amplifier, digital-to-analog, and analog-to-digital converters, is limited due to the high PAPR of the signals.
\par
An elegant solution to solve this power-control problem is to draw the modulating words from a block code that contains only words with low PAPR and, simultaneously, has error-correction capabilities. This idea was originally proposed for orthogonal frequency-division multiplexing (OFDM) systems \cite{Jones1996}, where a similar power-control problem occurs. We will see that the PAPR is always at least $1$. A code for which all codewords achieve this lower bound is called a \emph{constant-amplitude code}.
\par
A coding-theoretic framework for binary codes in MC-CDMA has been established by Paterson \cite{Paterson2004}. It was shown that codewords with low PAPR are exactly those words that are far from the first-order Reed--Muller code, $\RM(1,m)$. This fact was used in \cite{Paterson2004} to prove fundamental bounds on the trade-off between PAPR, minimum distance, and rate of binary codes. Moreover several families of binary constant-amplitude codes have been constructed in \cite{Paterson2004} by exploiting the relation between bent functions \cite{Rothaus1976}, \cite{MacWilliams1977} and binary constant-amplitude codewords (a connection that was first recognized by Wada \cite{Wada2000}). These codes are unions of cosets of $\RM(1,m)$ lying in higher-order Reed--Muller, Kerdock, or Delsarte--Goethals codes. Therefore they enjoy high minimum distance and are amenable to efficient encoding and decoding algorithms. 
\par
Sol\'{e} and Zinoviev \cite{Sole2006} constructed binary codes with PAPR much greater than $1$, which makes them less attractive for practical values of $n$. However in many situations their parameters asymptotically beat the Gilbert--Varshamov-style lower bound derived in \cite{Paterson2004}.
\par
While previous work was focused on binary codes, several motivations exist to study quaternary codes for MC-CDMA. First, quaternary modulation rather than binary modulation is often employed in MC-CDMA systems \cite{Dahlman1998}. Second, binary constant-amplitude codes cannot exist for lengths $2^m$ when $m$ is odd (we will see that quaternary codes indeed do). 
\par
In \cite{Schmidt2006c} the author established a connection between the coding problems in OFDM and in MC-CDMA. As a consequence, the generally nonbinary codes in \cite{Davis1999}, \cite{Paterson2000a}, \cite{Schmidt2007}, and \cite{Schmidt2006c} developed for OFDM can be re-used directly in MC-CDMA. However the best known upper bound on the PAPR of such codes is $2$, and it can be shown using the proof of \cite[Thm. 18]{Schmidt2006c} and the remarks thereafter that this approach cannot be used to construct quaternary constant-amplitude codes. 
\par
The key concept in this paper is the connection between words with PAPR equal to $1$ and functions from the binary $m$-tuples to $\Z_4$ having the bent property, i.e., the absolute values of their Fourier transform take on a constant value. This connection together with results from the theory of algebraic codes over $\Z_4$ is employed to construct families of quaternary constant-amplitude codes. Many of the resulting code families may be viewed as quaternary analogs of those developed in \cite{Paterson2004}.
\par
The remainder of this paper is organized as follows. In Section \ref{sec:system} we will state the coding problem in MC-CDMA formally. In Section~\ref{sec:preliminaries} we establish our main notation and prove some basic properties of quaternary constant-amplitude codes. In Section~\ref{sec:mappings}  mappings from binary to quaternary constant-amplitude codes are studied. In Sections~\ref{sec:Reed_Muller} and \ref{sec:K_DG} we construct constant-amplitude codes from quaternary Reed--Muller codes and from quaternary Kerdock and Delsarte--Goethals codes, respectively. Section~\ref{sec:conclusions} contains some final remarks.


\section{System Model and Problem Statement}
\label{sec:system}

We work with a simplified discrete-time MC-CDMA model, which essentially follows that in \cite{Paterson2004}. Let $n:=2^m$, and let the canonical Walsh--Hadamard matrix of size $n\times n$ be recursively defined by
\beqn
\label{eqn:Hadamard}
H_{n}:=\begin{pmatrix}
H_{n/2} & H_{n/2}\\
H_{n/2} & -H_{n/2}
\end{pmatrix}\quad
\mbox{with}\quad H_1:=(1).
\eeqn
Let $\omega$ be a primitive $2^h$th root of unity in $\Cn$, e.g., $\omega=\exp(i2\pi/2^h)$, where $i^2=-1$. Given a $\Z_{2^h}$-valued word $c=(c_0,c_1,\dots ,c_{n-1})$, the transmitted MC-CDMA signal can be modeled as
\beqn
\label{eqn:def_signal}
S_c(t)=\sum_{j=0}^{n-1}\omega^{c_j}(H_n)_{j,t},\quad t=0,1,\dots,n-1.
\eeqn
In words, $c_j$ is used to modulate the $j$th row of $H_n$, and the transmitted signal is the sum of these modulated sequences. The PAPR of the word $c$ (or of the signal $S_c(t)$) is defined as
\beqn
\label{eqn:def_PAPR}
\PAPR(c):=\frac{1}{n}\max_{0\le t<n}|S_c(t)|^2.
\eeqn
Observe that the PAPR is at least $1$ and can be as much as $n$. 
\par
A \emph{code} $\C$ of length $n$ over $\Z_{2^h}$ is defined as a subset $\C\subseteq\Z_{2^h}^n$. $\C$ is called \emph{linear} if it is a subgroup of $\Z_{2^h}^n$. We are particularly interested in the cases where $h=1$ or $h=2$, in which cases we say that $\C$ is \emph{binary} or \emph{quaternary}, respectively. If $\C$ is a binary code, then $\d_H(\C)$ denotes its minimum Hamming distance. Likewise, if $\C$ is a quaternary code, then $\d_L(\C)$ denotes its minimum Lee distance (see, e.g., \cite{Hammons1994} for details on Lee distance). The \emph{rate} of $\C$ is defined as
\beq
R(\C):=\frac{1}{n}\log_2|\C|,
\eeq
and the PAPR of $\C$ is defined to be
\beq
\PAPR(\C):=\max_{c\in\C}\PAPR(c).
\eeq
Our goal is to design quaternary codes $\C$ with $\PAPR(\C)$ being much lower than $n$. Ideally $\PAPR(\C)$ is equal to $1$, in which case $\C$ is called \emph{constant-amplitude code}.


\section{Generalized Bent Functions}
\label{sec:preliminaries}

Let 
\beq
F:=\{z\in\Z_{2^h}:z^2=z\}
\eeq
be the set of Teichmuller representatives in $\Z_{2^h}$. Then every $z\in\Z_{2^h}$ can be written uniquely in $2$-adic expansion
\beqn
\label{eqn:z_2adic}
z=\sum_{j=0}^{h-1}z_j2^j, \quad\mbox{where $z_0,z_1,\dots,z_{h-1}\in F$}.
\eeqn
We define an operation on $F$ by $a\oplus b:=(a+b)^{2^{h-1}}$.  Then $(F,\oplus,\cdot)\cong(\Z_2,+,\cdot)$ is the binary field. The operation `$+$' is reserved to denote addition in $\Z_{2^h}$.
\par
A \emph{generalized Boolean function} is defined as a mapping $f:F^m\rightarrow\Z_{2^h}$. Writing $k=(k_0,k_1,\dots,k_{m-1})$ for $k\in\{0,1\}^m$, every such function can be uniquely expressed in the polynomial form
\beq
f(x)=f(x_0,\dots,x_{m-1})=\sum_{k\in \{0,1\}^m}c_k\prod_{j=0}^{m-1} x_j^{k_j},\quad c_k\in\Z_{2^h},
\eeq
called the \emph{algebraic normal form} of $f$. The \emph{degree} of $f$ is defined to be 
\beq
\max_{c_k\ne 0}\;{\rm wt}_H(k),
\eeq
where ${\rm wt}_H(k)$ is the Hamming weight of $k$. Functions having algebraic normal form
\beq
f(x_0,\dots,x_{m-1})=x_0^{k_0}x_1^{k_1}\cdots x_{m-1}^{k_{m-1}},
\eeq
where $k_0,k_1,\dots,k_{m-1}\in \{0,1\}$, are called \emph{monomials}. Every generalized Boolean function can be expressed as a $\Z_{2^h}$-linear combination of monomials. By (\ref{eqn:z_2adic}), every $f:F^m\to\Z_{2^h}$ can be written uniquely in $2$-adic expansion, viz
\beqn
\label{eqn:f-2adic}
f(x_0,\dots,x_{m-1})=\sum_{j=0}^{h-1}f_j(x_0,\dots,x_{m-1})2^j,
\eeqn
where each $f_j$ is a mapping from $F^m$ to $F$.
\par
With each generalized Boolean function $f:F^m\rightarrow\Z_{2^h}$ we associate a word of length $2^m$ with elements in $\Z_{2^h}$. This word is obtained from the algebraic normal form of $f$ by listing all the values $f(x)$ as $x$ ranges over $F^m$ in lexicographic order. That is, if $\ell=\sum_{j=0}^{m-1}\ell_j2^j$, where $\ell_0,\ell_1,\dots,\ell_{m-1}\in F$, the $\ell$th element of this word reads $f(\ell_0,\ell_1,\dots,\ell_{m-1})$. By convention, we shall denote the function and the associated word by the same symbol, and since there is a one-to-one correspondence between them, we often use the terms \emph{function} and \emph{associated word} interchangeably.
\par
The \emph{$r$th-order Reed--Muller code} $\RM(r,m)$ of length $2^m$ is defined as the binary linear code that is generated by the $m$-variate monomials of degree at most $r$ \cite[Ch.~13]{MacWilliams1977}. It contains 
\beq
2^{\sum_{j=0}^r{m\choose j}}
\eeq
codewords and has minimum Hamming distance $2^{m-r}$. 
\par
The \emph{Fourier transform} of a function $f:F^m\to\Z_{2^h}$ is given by $\widehat f:
F^m\to\Cn$ with
\beq
\widehat f(u)=\sum_{x\in F^m}\omega^{f(x)}(-1)^{u\cdot x},
\eeq
where $\cdot$ denotes the scalar product in $F^m$ and $\omega$ is a primitive $2^h$th root of unity in $\Cn$. The multiset $\{\widehat f(u):u\in F^m\}$ is called the \emph{Fourier spectrum} of $f$.
\begin{definition}
A function $f:F^m\rightarrow\Z_{2^h}$ is \emph{bent} if $|\widehat f(u)|=2^{m/2}$ for every $u\in F^m$.
\end{definition}
\par
The name bent function was coined by Rothaus \cite{Rothaus1976}. His definition of bent functions applies to functions from $F^m$ to $\Z_2$. Kumar et al. \cite{Kumar1985} generalized this definition to functions from $\Z_q^m$ to $\Z_q$ for arbitrary $q$. Another generalization of bent functions appeared in \cite{Matsufuji1990}, where the functions are defined from $F^m$ to $\{0,1/2,1,3/2\}\cong\Z_4$ (as an additive group) and are called \emph{real-valued bent functions}. The latter essentially coincide with $\Z_4$-valued bent functions defined above.
\par
The following theorem will be the key to obtain our code constructions later in this paper. 
\begin{theorem}
\label{thm:PAPR-Fourier}
Let $c: F^m\rightarrow\Z_{2^h}$ be a generalized Boolean function. Then we have
\beq
\PAPR(c)=\frac{1}{2^m}\max_{u\in F^m}|\widehat c(u)|^2.
\eeq
In particular the PAPR of $c$ is equal to $1$ if and only if $c$ is a bent function.
\end{theorem}
\begin{proof}
Observe that the elements of the Walsh--Hadamard matrix $H_{2^m}$, shown in (\ref{eqn:Hadamard}), are given by
\beq
(H_{2^m})_{\ell,t}=(-1)^{\sum_{j=0}^{m-1}\ell_j t_j},
\eeq
where $\ell=\sum_{j=0}^{m-1}\ell_j2^j$ and $t=\sum_{j=0}^{m-1}t_j2^j$ with $\ell_j,t_j\in F$ for every $j=0,1,\dots,m-1$. Thus, for each generalized Boolean function $c:F^m\rightarrow\Z_{2^h}$, the corresponding MC-CDMA signal, given in (\ref{eqn:def_signal}), satisfies
\beq
S_c(t)=\widehat c(t_0,t_1,\dots,t_{m-1}).
\eeq
The first statement in the theorem then follows from (\ref{eqn:def_PAPR}). We immediately conclude that, $\PAPR(c)=1$ if $c$ is a bent function. The converse is a consequence of Parseval's identity 
\[
\frac{1}{2^m}\sum_{u\in F^m}|\widehat c(u)|^2=2^m.
\]
\vspace{-3ex}\par
\end{proof}
\par
We remark that in the binary case (where $h=1$), Theorem~\ref{thm:PAPR-Fourier} was essentially stated by Wada \cite{Wada2000} and by Paterson \cite[Thm.~8]{Paterson2004}.
\par
It is known (see, e.g., \cite{Rothaus1976}) that a $\Z_2$-valued bent function cannot exist if $m$ is odd and that the degree of a $\Z_2$-valued bent function is at most $m/2$ for $m>2$. Therefore binary constant-amplitude codes can only exist for even $m$ and must be contained in $\RM(m/2,m)$ if $m>2$. This gives us an upper bound on the size of a binary constant amplitude code and, perhaps surprisingly, a lower bound on its minimum Hamming distance. In the remainder of this section we shall derive analogous bounds for quaternary constant-amplitude codes.
\begin{lemma}
\label{lem:spectrum}
The values in the Fourier spectrum of a $\Z_{4}$-valued bent function on $F^m$ are of the form $2^{m/2}\omega^mi^k$, where $\omega=(1+i)/\sqrt{2}$ and $k\in \Z_4$.
\end{lemma}
\begin{proof}
Let $S$ denote an arbitrary value in the Fourier spectrum of a $\Z_4$-valued bent function on $F^m$. Then $\Re(S)$ and $\Im(S)$ must be integers and $|S|^2=2^m$ must be a sum of two squares. From Jacobi's two-square theorem we know that $2^m$ has a unique representation as a sum of two squares, namely $2^m=(2^{m/2})^2+0^2$ if $m$ is even, and $2^m=(2^{(m-1)/2})^2+(2^{(m-1)/2})^2$ if $m$ is odd. Hence, if $m$ is even, either $\Re(S)$ or $\Im(S)$ must be zero. If $m$ is odd, we must have $|\Re(S)|=|\Im(S)|$, which proves the lemma.
\end{proof}
\par
Recall from (\ref{eqn:f-2adic}) that every $f:F^m\to\Z_4$ can be written uniquely in $2$-adic expansion, viz
\beq
f(x)=a(x)+2b(x),
\eeq
where $a,b:F^m\to F$. We can express the Fourier transform of $f$ in terms of the Fourier transforms of $a$ and $b$ as follows
\begin{align}
\widehat f(u)&=\sum_{x\in F^m}i^{a(x)+2b(x)}(-1)^{u\cdot x}\nonumber\\
&=\frac{\widehat b(u)+(\widehat{a\oplus b})(u)}{2}+i\cdot\frac{\widehat b(u)-(\widehat{a\oplus b})(u)}{2},
\label{eqn:Fourier_exp}
\end{align}
which follows directly from the identity
\beq
i^{a+2b}=(-1)^{b}\cdot\frac{1+(-1)^a+i[1-(-1)^a]}{2},\quad a,b\in F.
\eeq
\begin{theorem}
\label{thm:bent_degree}
With the notation as above, let $m>2$ and suppose that $f:F^m\to\Z_4$ is a bent function. Then the degrees of $a$ and $b$ are at most $\lceil m/2\rceil$.
\end{theorem}
\begin{proof}
First suppose that $m$ is even. Then, by Lemma~\ref{lem:spectrum}, either the real part or the imaginary part of $\widehat f(u)$ must be zero. This can only happen if $b$ and $a\oplus b$ are bent, which implies that the degrees of $a$ and $b$ are at most $m/2$. Now let $m$ be odd. By Lemma~\ref{lem:spectrum}, in this case the real part and the imaginary part of $\widehat f(u)$ must be equal in magnitude. This is only possible if the Fourier transforms of $a\oplus b$ and $b$ take on only the values $0$ and $\pm 2^{(m+1)/2}$. Such functions are called \emph{almost bent functions} in \cite{Carlet1998a}. It is known \cite[Thm.~1]{Carlet1998a} that the degree of such a function is at most $(m+1)/2$.
\end{proof}
\par
\begin{theorem}
For $m>2$ every quaternary constant-amplitude code $\Q$ of length $2^m$ satisfies
\beq
R(\Q)<\frac{1}{2^{m-1}}\sum_{j=0}^{\lceil m/2\rceil}{m\choose j}.
\eeq
and
\beq
\d_L(\Q)\ge 2^{\lfloor m/2\rfloor}.
\eeq
\end{theorem}
\vspace{2ex}
\begin{proof}
By Theorem~\ref{thm:bent_degree}, every quaternary constant-amplitude code is contained inside the code
\beqn
\label{eqn:code_sum}
\{a+2b:a,b\in \RM(\lceil m/2\rceil,m)\},
\eeqn
whose minimum Lee distance is at least the minimum Hamming distance of $\RM(\lceil m/2\rceil,m)$. Since at least one word in the code (\ref{eqn:code_sum}) has PAPR greater than one (e.g., the all-zero codeword), it cannot be itself a constant-amplitude code. The theorem follows from the properties of $\RM(\lceil m/2\rceil,m)$.
\end{proof}
\par
The preceding theorem provides a reasonable rate bound for small $m$. For instance, if $m=4$, the bound asserts that the rate of a quaternary constant-amplitude is at most $21/16$ (assuming an integer number of encoded bits). We will later construct a code with rate $18/16$ (see Table~\ref{tab:ca-codes}). We speculate that the rate bound tends to be loose for large $m$. However, even for large $m$, the bound says that the constant-amplitude property of a quaternary code must be paid by a rate reduction from $2$ to at most slightly more than $1$.


\section{Mappings From Binary to Quaternary Codes}
\label{sec:mappings}

In this section we study mappings from binary to quaternary constant-amplitude codes, which can be used in connection with the results obtained in \cite[Sec.~V]{Paterson2004} to construct quaternary constant-amplitude codes. Recall from Theorem~\ref{thm:PAPR-Fourier} that every constant-amplitude code must be comprised of bent functions. Therefore we shall study mappings from $\Z_2$-valued bent functions to $\Z_4$-valued bent functions.
\par
Some mappings are based on the \emph{Gray map} $\phi:\Z_4\to F^2$ given by
\beq
0 \mapsto (00),\quad 1 \mapsto (01),\quad 2 \mapsto (11),\quad 3 \mapsto (10).
\eeq 
In other words, writing $z=a+2b$ where $a,b\in F$, we have $\phi(z)=(b,a\oplus b)$. We extend $\phi$ naturally to act on words in $\Z_4^n$. A useful property of $\phi$ is that it is a distance-preserving bijection from $\Z_4^n$ equipped with the Lee distance to $F^{2n}$ equipped with the Hamming distance \cite{Hammons1994}. That is, if $\Q$ is a quaternary code, we have
\beq
\d_L(\Q)=\d_H(\phi(\Q)).
\eeq
If $f$ is a $\Z_4$-valued word of length $2^m$, we will often be interested in the $\Z_2$-valued Boolean function that describes $\phi(f)$. If we write $f(x)$ in $2$-adic expansion, viz $f(x)=a(x)+2b(x)$ where $a,b$ are $F$-valued, then by the definition of the Gray map, $\phi(f)$ is a function from $F^{m+1}$ to $F$ given by $\phi(f)(x,y)=a(x)y\oplus b(x)$.
\par
The theorem below has been essentially stated in \cite[Thm.~2]{Novosad1993}.
\begin{theorem}
\label{thm:bent_map_m_odd_1}
Let $m\ge 1$, and let $a,b:F^{m-1}\rightarrow F$ be bent functions. Then $f:F^m\rightarrow\Z_4$ given by
\beq
f(x,y)=2a(x)(1+y)+2b(x) y+y
\eeq
is also bent.
\end{theorem}
\begin{proof} Similar to the proof  of \cite[Thm.~2]{Novosad1993}.
\end{proof}
\begin{corollary}
\label{cor:code_map_m_odd_1}
Let $\B$ be a binary constant-amplitude code of length $2^{m-1}$ (so $m$ is necessarily odd). Then the code
\beq
\Q=\{(2a+\epsilon\cdot \bone\,,\,2b+(1+\epsilon)\cdot\bone):a,b\in\B,\epsilon\in F\}
\eeq 
of length $2^m$ is a quaternary constant-amplitude code with
\begin{align*}
\d_L(\Q)&=2\d_H(\B)\\
R(\Q)&=R(\B)+1/2^m.
\end{align*}
Here, $\bone$ denotes the all-one word of length $2^{m-1}$.
\end{corollary}
\par
Note that the quaternary code in the preceding corollary is essentially an offset binary code that is obtained by concatenating two suitably offset codewords of a binary constant-amplitude code. This has to be compared with \cite[Lem.~19 and the remarks thereafter]{Paterson2004}, where essentially the same code construction is suggested resulting in, however, codes with PAPR equal to $2$.
\par
\begin{theorem}
\label{thm:bent_map_m_even}
Let $a,b:F^m\rightarrow F$ be bent functions, and define $g:F^{m+1}\rightarrow F$ by
\beq
g(x,y)=a(x)y\oplus b(x)(1\oplus y).
\eeq
Then $f=\phi^{-1}(g)$ is a bent function. 
\end{theorem}
\begin{proof}
By rewriting $g(x,y)$ as
\beq
g(x,y)=b(x)\oplus [a(x)\oplus b(x)]y,
\eeq
it is seen that $f(x)$ has $2$-adic expansion
\beq
f(x)=2b(x)+[a(x)\oplus b(x)].
\eeq
Using (\ref{eqn:Fourier_exp}), we obtain
\beq
\widehat f(u)=\frac{\widehat b(u)+\widehat a(u)}{2}+i\cdot\frac{\widehat b(u)-\widehat a(u)}{2}.
\eeq
Since $ \widehat a(u)$ and $ \widehat b(u)$ take on only the values $\pm 2^{m/2}$, it follows that $|\widehat f(u)|=2^{m/2}$ for each $u\in F^m$.
\end{proof}
\par
The preceding theorem together with the distance-preserving property of the Gray map implies the following.
\begin{corollary}
\label{cor:code_map_m_even}
Let $\B$ be a binary constant-amplitude code of length $2^m$ (so $m$ is necessarily even). Then the quaternary code
\beq
\Q=\phi^{-1}\left(\{(a,b):a,b\in\B\}\right)
\eeq 
of length $2^m$ is a quaternary constant-amplitude code with 
\begin{align*}
\d_L(\Q)&=\d_H(\B)\\
R(\Q)&=2R(\B).
\end{align*}
\end{corollary}
\vspace{2ex}
\par
\begin{theorem}
\label{thm:bent_map_m_odd_2}
Let $m\ge 1$, and let $g:F^{m+1}\rightarrow F$ be a bent function. Then $f=\phi^{-1}(g)$ is also bent.
\end{theorem}
\begin{proof}
Recall that, when $g$ is expressed as $g(x,y)=a(x)\,y\oplus b(x)$, where $a,b:F^m\rightarrow F$, then $f$ is given by $f(x)=a(x)+2b(x)$. The Fourier transform of $g$ can be written as
\begin{align*}
\widehat g(u,v)&=\sum_{x\in F^m,y\in F}(-1)^{a(x)y \oplus b(x) \oplus u\cdot x \oplus vy}\\
&=\sum_{x\in F^m}(-1)^{b(x) \oplus u\cdot x}(1+(-1)^{v\oplus a(x)}).
\end{align*}
Using the expansion (\ref{eqn:Fourier_exp}) we obtain
\beq
\widehat f(u)=\frac{\widehat g(u,0)}{2}+i\cdot\frac{\widehat g(u,1)}{2}.
\eeq
Since, by assumption, $|\widehat g(u,v)|=2^{(m+1)/2}$ for each $u\in F^m$ and each $v\in F$, we have $|\widehat f(u)|=2^{m/2}$ for each $u\in F^m$.
\end{proof}
\par
With the distance-preserving property of the Gray map we obtain our next corollary.
\begin{corollary}
\label{cor:code_map_m_odd_2}
Let $\B$ be a binary constant-amplitude code of length $2^{m+1}$ (so $m$ must be odd). Then $\Q=\phi^{-1}(\B)$ is a quaternary constant-amplitude code of length $2^{m}$ with
\begin{align*}
\d_L(\Q)&=\d_H(\B)\\
R(\Q)&=2R(\B).
\end{align*}
\end{corollary}


\section{Codes From Quaternary Reed--Muller Codes}
\label{sec:Reed_Muller}

Below we recall two quaternary generalizations of the binary Reed--Muller code from \cite{Davis1999}.\pagebreak[4]
\begin{definition}~
\begin{enumerate}[a)]
\item For $0\le r\le m$ the code $\RM_4(r,m)$ is defined as a linear code over $\Z_4$ of length $2^m$ that is generated by the $m$-variate monomials of degree at most $r$ \cite{Davis1999}.
\item For $0\le r\le m+1$ the code $\ZRM(r,m)$ is defined as a linear code over $\Z_4$ of length $2^m$ that is generated by the $m$-variate monomials of degree at most $r-1$ together with two times the $m$-variate monomials of degree $r$ with the convention that the monomials of degree $-1$ and $m+1$ are equal to zero \cite{Davis1999} (see also \cite{Hammons1994} for a slightly different, generally nonequivalent, definition of $\ZRM(r,m)$).
\end{enumerate}
\end{definition}
\par
The code $\RM_4(r,m)$ contains 
\beq
4^{\sum_{j=0}^{r}{m\choose j}}
\eeq 
codewords and has minimum Lee distance $2^{m-r}$, while $\ZRM(r,m)$ is a subcode of $\RM_4(r,m)$ that contains 
\beq
4^{\sum_{j=0}^{r-1}{m\choose j}}\cdot2^{{m\choose r}}
\eeq
codewords and has minimum Lee distance $2^{m-r+1}$ \cite{Davis1999}.
\par
The next lemma will be useful in the sequel.
\begin{lemma}
\label{lem:PAPR_ZRM}
Every word in a coset of $\ZRM(1,m)$ has the same PAPR.
\end{lemma}
\begin{proof}
The lemma is a direct consequence of the fact that the absolute values in the Fourier spectrum of $f:F^m\rightarrow\Z_4$ and the absolute values in the Fourier spectrum of a function given by
\beq
f(x)+2c\cdot x+\epsilon,\quad c\in F^m,\epsilon\in\Z_4
\eeq
are identical.
\end{proof}
\par
The following theorem identifies a large number of $\Z_4$-valued bent functions.
\begin{theorem}
\label{thm:MF}
Let $\sigma $ be a permutation on $F^k$, let $g:F^k\rightarrow\Z_4$ be arbitrary, and let $f:F^{2k}\rightarrow\Z_4$ be given by
\beq
f(x,y)=2\,\sigma(x)\cdot y+g(x).
\eeq
Then $f$ is a bent function.
\end{theorem}
\begin{proof}
Write
\begin{align*}
\widehat f(u,v)&=\sum_{x,y\in F^k}i^{2\sigma(x)\cdot y+g(x)}(-1)^{u\cdot x+v\cdot y}\\
&=\sum_{x\in F^k}i^{g(x)}(-1)^{u\cdot x}\cdot\sum_{y\in F^k}(-1)^{(\sigma(x)+v)\cdot y}\\
&=\sum_{x\in F^k}i^{g(\sigma^{-1}(x))}(-1)^{u\cdot \sigma^{-1}(x)}\cdot\sum_{y\in F^k}(-1)^{(x+v)\cdot y}.
\end{align*}
If $x\ne v$, the exponent in the inner sum is a linear form in $y$. Hence the inner sum is zero if $x\ne v$. On the other hand, if $x=v$, the sum is trivial and becomes equal to $2^k$. Thus
\beq
\widehat f(u,v)=2^k\cdot i^{g(\sigma^{-1}(v))}(-1)^{u\cdot \sigma^{-1}(v)},
\eeq
and the proof is completed.
\end{proof}
\par
The preceding theorem could be understood as a generalization of the Maiorana--McFarland construction of $\Z_2$-valued bent functions (see, e.g., \cite[Ch. 14, Problem (20)]{MacWilliams1977}). A special case of this construction has been stated in \cite[Thm.~1]{Matsufuji1990} in terms of real-valued bent functions.
\par
Now write the permutation $\sigma$ in Theorem~\ref{thm:MF} as $(\sigma_0,\sigma_1,\dots,\sigma_{k-1})$, where $\sigma_0,\dots,\sigma_{k-1}:F^k\to F$. Every $\sigma_j$ must be balanced, so has degree at most $k-1$ (see, e.g., \cite[Ch. 13, Thm. 1]{MacWilliams1977}). Therefore the functions $f$ identified in Theorem~\ref{thm:MF} have degree at most $k$. We deduce the following coding option.
\par
\begin{construction}
\label{con:MF}
For $m=2k$ Theorem~\ref{thm:MF} identifies $2^{m/2}!\cdot4^{2^{m/2}}$ bent functions, which organize in cosets of $\ZRM(1,m)$ inside $\RM_4(m/2,m)$. The union of them is a constant-amplitude code of length $2^m$ with minimum Lee distance $2^{m/2}$. Assuming an integer number of encoded bits, for $m=4,6,8$ the rate of this code is equal to $12/16$, $31/64$, and $76/256$, respectively. By restricting the functions $g$ in Theorem~\ref{thm:MF} such that the coefficient of the monomial of degree $m/2$ in the algebraic normal form of $g$ is even, we can construct a subcode of the code considered above that is contained in $\ZRM(m/2,m)$. This code has half as many codewords as the code contained in $\RM_4(m/2,m)$, but minimum Lee distance $2^{m/2+1}$. The rate of this code is equal to $11/16$, $30/64$, and $75/256$ for $m=4,6,8$, respectively.
\end{construction}
\par
Now we turn our attention to the code $\ZRM(2,m)$. This code comprises $2^{m(m+1)/2}$ cosets of $\ZRM(1,m)$, where the coset representatives are given by a function $Q:F^m\to\Z_4$ that can be written as
\begin{align}
\label{eqn:quad_form}
Q(x)&=xBx^T\\
&=\sum_{j=0}^{m-1}b_{jj}x_j^2+2\sum_{0\le j<k<m}b_{jk}x_jx_k,\nonumber
\end{align}
where $B=(b_{jk})$ is an $F$-valued symmetric matrix of size $m\times m$. Such a $Q$ is called a \emph{$\Z_4$-valued quadratic form} \cite{Brown1972}. We say that $Q$ has rank $r$ and write $\rk(Q)=r$ if the matrix $B$ has rank $r$ (being computed over $F$). 
\begin{theorem}
\label{thm:rank_fourier}
With the notation as above,
\beq
|\widehat Q(u)|\le 2^{m-\rk(Q)/2}.
\eeq
\end{theorem}
\vspace{1ex}
\begin{proof}
We have
\begin{align*}
|\widehat Q(u)|^2&=\bigg(\sum_{x\in F^m}i^{xBx^T+2u\cdot x}\bigg)\bigg(\sum_{y\in F^m}i^{-yBy^T+2u\cdot y}\bigg)\\
&=\sum_{x,y\in F^m}i^{xBx^T-yBy^T+2u\cdot(x+y)}\\
&=\sum_{x,y\in F^m}i^{(x+y)B(x-y)^T+2u\cdot(x+y)}.
\end{align*}
Next we apply the variable substitution
\beq
z:=x\oplus y=x+y+2x*y,
\eeq
where $x*y$ denotes the component-wise product of $x$ and $y$. Notice that for each fixed $y$, the mapping $x\mapsto z$ is a bijection on $F^m$. Using the identities
\begin{align*}
x+y&=z-2x*y\\
x-y&=z-2x*y-2y,
\end{align*}
we can write
\begin{align*}
|\widehat Q(u)|^2&=\sum_{z,y\in F^m}i^{(z+2x*y)B(z+2x*y+2y)^T+2u\cdot z}\\
&=\sum_{z,y\in F^m}i^{zBz^T+2zBy^T+2u\cdot z}\\
&=\sum_{z\in F^m}i^{zBz^T+2u\cdot z}\sum_{y\in F^m}(-1)^{zBy^T}.
\end{align*}
If the rank of $B$ is $r$, then there exist $2^{m-r}$ elements $z\in F^m$ such that $zBy^T=0\pmod 2$ for each $y\in F^m$ and the inner sum is equal to $2^m$.  On the other hand, for the remaining $2^r$ elements $z\in F^m$, the expression  $zBy^T$ is a nonzero linear form in $y$ and the inner sum is zero.  We conclude that 
\beq
|\widehat Q(u)|^2\le 2^m\times 2^{m-r},
\eeq
which proves the theorem. 
\end{proof}
\par
The preceding theorem in connection with Theorem~\ref{thm:PAPR-Fourier} and Lemma~\ref{lem:PAPR_ZRM} yields an upper bound on the PAPR of the cosets of $\ZRM(1,m)$ inside $\ZRM(2,m)$.
\begin{corollary}
\label{cor:PAPR_coset}
Let $Q:F^m\to\Z_4$ be a $\Z_4$-valued quadratic form. Then the PAPR of the coset $Q+\ZRM(1,m)$ is at most $2^{m-\rk(Q)}$.
\end{corollary}
\par
A simple constant-amplitude code can now be constructed as follows.
\begin{construction}
\label{con:single_coset}
Let $Q:F^m\to\Z_4$ be a $\Z_4$-valued quadratic form of rank $m$. For instance, we may take
\beq
Q(x)=x_0+x_1+\cdots+x_{m-1}
\eeq
(in which case the corresponding $B$ is the identity matrix). Then the coset $Q+\ZRM(1,m)$ is a constant-amplitude code that contains $2^{m+2}$ codewords and has minimum Lee distance $2^m$.
\end{construction}
\par
In what follows we will construct a constant-amplitude subcode of $\ZRM(2,m)$ by collecting all cosets of  $\ZRM(1,m)$ inside $\ZRM(2,m)$ whose coset representatives correspond to $\Z_4$-valued quadratic forms of rank $m$. The number of such forms is equal to the number of $m\times m$ nonsingular symmetric matrices over $F$. This number is well known (see, e.g., \cite{MacWilliams1969}) and given by
\beq
\displaystyle
N(m)=\begin{cases}
\displaystyle
\prod_{j=1}^{m/2}(2^{m+1}-2^{2j})&\mbox{if}\;m\;\mbox{is even}\\
\displaystyle
\prod_{j=0}^{(m-1)/2}(2^m-2^{2j})&\mbox{if}\;m\;\mbox{is odd.}
\end{cases}
\eeq
\par
\begin{construction}
\label{con:zrm2}
$\ZRM(2,m)$ contains $N(m)$ cosets of $\ZRM(1,m)$ with PAPR equal to $1$. The union of $2^{\lfloor \log_2 N(m)\rfloor}$ such cosets is a constant-amplitude code of length $2^m$ with minimum Lee distance $2^{m-1}$ and rate
\beq
\frac{\lfloor\log_2N(m)\rfloor+m+2}{2^m}.
\eeq
For $m=4,5,6,7$ the rate is equal to $14/16$, $20/32$, $27/64$, $35/128$, respectively. A crude estimate yields $N(m)\ge 2^{\lfloor m^2/2\rfloor}$. Thus the rate of this code is bounded below by 
\beq
\frac{\lfloor m^2/2\rfloor+m+2}{2^m}.
\eeq
\end{construction}


\section{Codes From Quaternary Kerdock and Delsarte--Goethals Codes}
\label{sec:K_DG}

We will begin this section with a short review of symmetric bilinear forms, which we will use to construct the quaternary Kerdock and Delsarte--Goethals codes. These constructions are then utilized to derive two classes of quaternary constant-amplitude codes.
\par
We let $E$ be an extension field of $F$, so that $(E,\oplus,\cdot)$ is isomorphic to ${\rm GF}(2^m)$. A \emph{symmetric bilinear form} on $E$ is a mapping $\B:E\times E\to F$ that satisfies symmetry, i.e., 
\beq
\B(x,y)=\B(y,x)
\eeq
and the bilinearity condition
\beq
\B(x,ay\oplus bz)=a\B(x,y)\oplus b\B(x,z),\;\;\forall a,b\in F.
\eeq
Note that bilinearity in the first argument follows from symmetry. The \emph{radical} $\rad(\B)$ of $\B$ contains all elements $x\in E$ such that $\B(x,y)=0$ for all $y\in E$. It is a consequence of the bilinearity condition that $\rad(\B)$ is a subspace of $E$. The \emph{rank} of $\B$ is defined as the codimension of the radical, i.e., 
\beq
\rk(\B):=m-\dim\rad(\B).
\eeq
\par
Now let $\{\lambda_0,\lambda_1,\dots,\lambda_{m-1}\}$ be a fixed basis for $E$ over $F$. By the bilinearity condition we have
\beq
\B\left(\bigoplus_{j=0}^{m-1}x_j\lambda_j,\bigoplus_{k=0}^{m-1}y_k\lambda_k\right)=\bigoplus_{j=0}^{m-1}\bigoplus_{k=0}^{m-1}x_j\B(\lambda_j,\lambda_k)y_k
\eeq
for all $x_0,y_0,\dots,x_{m-1},y_{m-1}\in F$. Therefore, relative to the fixed basis for $E$ over $F$, the symmetric bilinear form $\B$ is completely determined by the $m\times m$ symmetric matrix $B$ whose elements at positions $(j,k)$ are given by $\B(\lambda_j,\lambda_k)$. Hence we can write
\beq
\B(x,y)=(x_0,\dots,x_{m-1})B(y_0,\dots,y_{m-1})^T,
\eeq
where $x=\bigoplus_{j=0}^{m-1}\lambda_jx_j$ and $y=\bigoplus_{j=0}^{m-1}\lambda_jy_j$. Note that the rank of the matrix $B$ is equal to $\rk(\B)$.
\par
For the remainder of this section we define the linearized polynomial
\beq
L_a(x):=\bigoplus_{j=1}^{t}(a_jx^{2^j}\oplus a_j^{2^{m-j}}x^{2^{m-j}})\oplus a_0x,
\eeq
where $a=(a_0,a_1,\dots,a_t)\in E^{t+1}$ and $0\le t<m/2$. We also define the mapping $\B_a:E\times E\to F$ by
\beq
\B_a(x,y):=\tr(yL_a(x)),
\eeq
where
\beq
\tr(x):=\bigoplus_{j=0}^{m-1}x^{2^j}
\eeq
is the trace function from $E$ to $F$. It is not hard to verify that $\B_a$ obeys the conditions of a symmetric bilinear form. We define $\M(t,m)$ to be the set of symmetric matrices that correspond to the set of bilinear forms $\{\B_a:a\in E^{t+1}\}$. The crucial property of $\M(t,m)$ is given by the following theorem.
\begin{theorem}
\label{thm:rank_M}
All nonzero matrices in $\M(t,m)$ have rank at least $m-2t$.
\end{theorem}
\begin{proof}
Since $\B_0$ corresponds to the all-zero matrix, we assume $a\ne 0$. Note that $\B_a(x,y)=0$ for every $y\in E$ if and only if $L_a(x)=0$. Now observe that $x\mapsto x^{2^t}$ is an automorphism on $E$ and 
\beq
L_a(x^{2^t})=\bigoplus_{j=1}^{t}(a_jx^{2^{t+j}}\oplus a_j^{2^{m-j}}x^{2^{t-j}})\oplus a_0x^{2^t}
\eeq
has degree at most $2^{2t}$. Therefore, $L_a(x)$ has at most $2^{2t}$ roots in $E$. We conclude that the number of $x\in E$ such that $\B_a(x,y)=0$ for every $y\in E$ is at most $2^{2t}$. Hence we have $\dim\rad(\B_a)\le 2t$, so the rank of $\B_a$ (and therefore the rank of the corresponding matrix) is at least $m-2t$.
\end{proof}
\par
Next we use the set $\M(t,m)$ to construct the quaternary Delsarte--Goethals code.
\begin{definition}
Let $\Q$ denote the set of $\Z_4$-valued quadratic forms corresponding to the symmetric matrices in $\M(t,m)$ (cf. (\ref{eqn:quad_form})). For $0\le t<m/2$ the \emph{quaternary Delsarte--Goethals code} is defined to be
\beq
\DG(t,m):=\bigcup_{Q\in\Q}Q+\ZRM(1,m).
\eeq
Moreover let $\K(m):=\DG(0,m)$ be the \emph{quaternary Kerdock code}.
\end{definition}
\par
Our definition of $\DG(t,m)$ essentially coincides with the definition given in \cite{Hammons1994}, which we have extended from odd $m$ to arbitrary $m$. We note that a similar definition of the quaternary Kerdock code also appears in \cite{Calderbank1997a}. 
\par
The code $\DG(t,m)$ is a linear subcode of $\ZRM(2,m)$ (because $\M(t,m)$ is closed under addition) and a union of $2^{m(t+1)}$ cosets of $\ZRM(1,m)$, so we have
\beq
|\DG(t,m)|=2^{m(t+2)+2}.
\eeq
It is known (see, e.g., \cite{Helleseth1996}) that
\beq
\d_L(\DG(t,m))=2^m-2^{t+\lfloor m/2\rfloor}.
\eeq
By Theorem~\ref{thm:rank_M} and Corollary~\ref{cor:PAPR_coset}, we have the following.
\begin{corollary}
The code $\DG(t,m)\setminus\ZRM(1,m)$ has PAPR at most $4^t$. 
\end{corollary}
\par
In the particular case where $t=0$ we obtain a constant-amplitude code.
\begin{construction}
\label{con:kerdock}
The code $\K(m)$ consists of $2^m$ cosets of $\ZRM(1,m)$, one of them is $\ZRM(1,m)$ itself. The union of $2^{m-1}$ remaining cosets is a constant-amplitude code of length $2^m$ with rate $(2m+1)/2^m$ and minimum Lee distance $2^m-2^{\lfloor m/2\rfloor}$.
\end{construction}
\par
In the remainder of this section we shall construct a constant-amplitude subcode of $\DG(1,m)$. The key to obtain the construction is given by the theorem below.
\begin{theorem}
\label{thm:M1}
The number of nonsingular matrices in $\M(1,m)$ is at least $4^{m-1}$.
\end{theorem}
\begin{proof}
For $t=1$ we have
\beq
L_a(x)=a_1x^2\oplus a_0x\oplus a_1^{2^{m-1}}x^{2^{m-1}}.
\eeq
We are interested in the number of cases where the equation $L_a(x)=0$ has only the trivial solution $x=0$. Squaring both sides and dividing both sides by $x$ yields
\beqn
\label{eqn:cubic}
a_1^2x^3\oplus a_0^2x\oplus a_1=0.
\eeqn
If $a_1=0$, there are $2^m-1$ choices for $a_0$ such that $L_a(x)=0$ has exactly one solution, so assume in the following that $a_1\ne 0$. There are two cases. If $a_0=0$, (\ref{eqn:cubic}) is equivalent to $x^3=1/a_1$. This equation has no solution if and only if $a_1^{-1}\not\in\{x^3:x\in E\}$, which can happen in
\beq
(2^m-1)\left(1-\frac{1}{\gcd(3,2^m-1)}\right)
\eeq
cases. The latter expression is zero if $m$ is odd and is equal to $\frac{2}{3}(2^m-1)$ if $m$ is even. If $a_0\ne 0$, (\ref{eqn:cubic}) becomes under variable substitution $x\mapsto \frac{a_0}{a_1}x$
\beq
x^3+x+\frac{a_1^2}{a_0^3}=0.
\eeq
From \cite[Appendix]{Kumar1996} we know that for each $a_0\in E\setminus\{0\}$ there exist $\frac{1}{3}(2^m-(-1)^m)$ different $a_1\in E\setminus\{0\}$ such that this equation has no solution.
\par
Straightforward counting yields that for either $m$ the set $\M(1,m)$ contains
\beq
2^m-1+\frac{(2^m+1)(2^m-1)}{3}>4^{m-1}
\eeq
nonsingular matrices, which completes the proof.
\end{proof}
\par
The preceding theorem leads to the following coding option.
\begin{construction}
\label{con:dg}
By Theorem~\ref{thm:M1}, $\DG(1,m)$ contains $2^{2m-2}$ cosets of $\ZRM(1,m)$ that correspond to $\Z_4$-valued quadratic forms of rank $m$. The union of these cosets is a constant-amplitude code of length $2^m$, rate $3m/2^m$, and minimum Lee distance $2^m-2^{1+\lfloor m/2\rfloor}$.
\end{construction}
\par
\begin{proofnote}
Recently, the author determined the complete rank distribution of the set $\M(t,m)$ \cite{Schmidt2008}, which can be used to show that the number of nonsingular matrices in $\M(t,m)$ is at least $2^{(t+1)m-2}$. Therefore, for general $t$, the code $\DG(t,m)$ contains $2^{(t+1)m-2}$ cosets of $\ZRM(1,m)$ that correspond to $\Z_4$-valued quadratic forms of rank $m$. The union of these cosets is a constant-amplitude code of length $2^m$, rate $(t+2)m/2^m$, and minimum Lee distance $2^m-2^{t+\lfloor m/2\rfloor}$.
\end{proofnote}


\section{Final Remarks}
\label{sec:conclusions}

We have studied quaternary codes for PAPR reduction in MC-CDMA for the first time. Several constructions of codes with PAPR equal to $1$ have been established. These codes are unions of cosets of $\ZRM(1,m)$. It is not hard to show that the image under the Gray map of such a code is a union of cosets of $\RM(1,m+1)$. Since encoders and decoders for the binary image may be used to encode and decode, respectively, the corresponding quaternary code, existing encoding and decoding algorithms developed for binary OFDM codes \cite{Davis1999}, \cite{Paterson2000b} are directly applicable here.
\par
For $m\in\{4,5,6\}$ the parameters of several constant-amplitude codes constructed in this paper are shown in Table~\ref{tab:ca-codes}. This table illustrates how code rate can be traded against minimum Lee distance. Note that we have omitted those codes that were outperformed by one of the codes shown in the table. In particular the codes flowing from Construction \ref{con:MF} are notoriously surpassed by other codes; they appear to be rather weak at least for small $m$.
\setlength{\tabcolsep}{20pt}
\begin{table*}[t]
\centering
\caption{Parameters of Quaternary Constant-Amplitude Codes}
\label{tab:ca-codes}
\begin{tabular}{c|c|c|c}
\hline
$m$ & rate & min. Lee dist. & reference\\\hline\hline
4 & 6/16 & 16        & Construction~\ref{con:single_coset},      single coset of $\ZRM(1,4)$\\
  & 9/16 & 12        & Construction~\ref{con:kerdock}, subcode of $\K(4)$\\
  & 14/16 & 8        & Construction~\ref{con:zrm2}, subcode of $\ZRM(2,4)$\\
  & 18/16 & 4        & \cite[Construction 12]{Paterson2004} + Corollary~\ref{cor:code_map_m_even}\\ \hline
  
5 & 7/32 & 32        & Construction~\ref{con:single_coset}, single coset of $\ZRM(1,5)$\\
  & 11/32 & 28       & Construction~\ref{con:kerdock}, subcode of $\K(5)$\\
  & 15/32 & 24       & Construction~\ref{con:dg}, subcode of $\DG(1,5)$\\
  & 20/32 & 16       & Construction~\ref{con:zrm2}, subcode of $\ZRM(2,5)$\\
  & 23/32 & 8        & \cite[Construction 13]{Paterson2004} + Corollary~\ref{cor:code_map_m_odd_2}\\\hline
  
6 & 8/64 & 64        & Construction~\ref{con:single_coset}, single coset of $\ZRM(1,6)$\\
  & 13/64 & 56       & Construction~\ref{con:kerdock}, subcode of $\K(6)$\\
  & 18/64 & 48       & Construction~\ref{con:dg}, subcode of $\DG(1,6)$\\
  & 27/64 & 32       & Construction~\ref{con:zrm2}, subcode of $\ZRM(2,6)$\\
  & 30/64 & 24       & \cite[Construction 18]{Paterson2004} + Corollary~\ref{cor:code_map_m_even}\\
  & 40/64 & 16       & \cite[Construction 12]{Paterson2004} + Corollary~\ref{cor:code_map_m_even}\\
  & 46/64 & 8        & \cite[Construction 13]{Paterson2004} + Corollary~\ref{cor:code_map_m_even}\\ \hline

\end{tabular}
\end{table*}
\par
Finally we wish to compare code rates and minimum Euclidean distances of our quaternary codes with those of previously constructed binary codes. Recall that the minimum Euclidean distance of a binary (quaternary) code is equal to four (two) times its minimum Hamming (Lee) distance. It can be observed that all binary constant-amplitude codes proposed in \cite[Sec.~V]{Paterson2004} are outperformed by a quaternary constant-amplitude code constructed in the present paper. For example, \cite{Paterson2004} reports a code of length $64$ with minimum Euclidean distance $32$ and rate $23/64$, while Table~\ref{tab:ca-codes} contains a quaternary code with the same length and minimum Euclidean distance but rate $40/64$.



\end{document}